\documentclass[ 
reprint,
superscriptaddress,
 amsmath,amssymb,
 aps,
floatfix,
]{revtex4-2}

\usepackage{physics}
\usepackage{graphicx}
\usepackage{dcolumn}
\usepackage{bm}
\usepackage{xcolor}

\begin{document}

\preprint{APS/123-QED}

\title{Hierarchical localization in disordered Apollonian networks}

\author{Eduardo M. K. Souza}
\affiliation{
 Instituto de Física, Universidade Federal de Alagoas, 57072-900 Maceió, Alagoas, Brazil
}


\author{Francisco A. B. F. de Moura}
\affiliation{
 Instituto de Física, Universidade Federal de Alagoas, 57072-900 Maceió, Alagoas, Brazil
}

\author{Guilherme M. A. Almeida}%
\affiliation{
 Instituto de Física, Universidade Federal de Alagoas, 57072-900 Maceió, Alagoas, Brazil
}

\date{\today}

\begin{abstract}
We investigate localization properties of the Apollonian network (AN) in the presence of diagonal and off-diagonal disorder. By employing a site-resolved localization measure, we show that the localization degree is strongly dependent on the energy and 
tied to the hierarchical topology of the network.
At the spectral edges, eigenstates are strongly localized on highly connected sites originating from previous generations, 
a behavior that persists under both disorder mechanisms. In contrast, around zero energy localization is associated with the lowest-degree sites. As disorder breaks the underlying 
$C3$ symmetry of the AN, it promotes spatial reconfiguration of these states while preserving their support on low-degree nodes.
For diagonal disorder, localization is enhanced over a broad range of negative energies, whereas off-diagonal disorder induces weakening of localization in this region. Finally, we show that the hub dominates the spectral edges but has negligible contribution near the band center, indicating that its associated localized states are robust against disorder. These results highlight how topology and disorder jointly shape localization in complex networks.
%

%
\end{abstract}

\maketitle


\section{\label{sec:level1} INTRODUCTION}


Network theory is a powerful tool for understanding the structure and dynamics of a wide variety of complex systems found in nature and technology \cite{albert02,boccaletti06}.
Examples include disease spreading \cite{pandemic,pandemic2}, 
transportation networks \cite{verma14,meu3}, 
quantum technologies \cite{almeida13,zhuang21,brito21,nokkala24}, 
condensed matter physics \cite{andrade05mag,souza07,oliveira10,CondMatter2},
and more. 

While many network models are employed to describe specific scenarios, real-world networks often share patterns characterized by hierarchical and community structures, power-law degree distribution, small-world behavior, and other non-trivial features.
Paradigmatic network models such as Watts-Strogatz \cite{SmallWorld}, Barabási-Albert \cite{Scale}, and Erdös-Rényi \cite{Erdos} are part of a broader class known as complex networks.
%
Among these, Apollonian networks (ANs) \cite{prlApolo} stand out due to their unique combination of such properties, 
which also includes self-similarity, 
all emerging from a simple deterministic construction.

\begin{figure}[t!]
    \centering
    \includegraphics[width=0.9\linewidth]{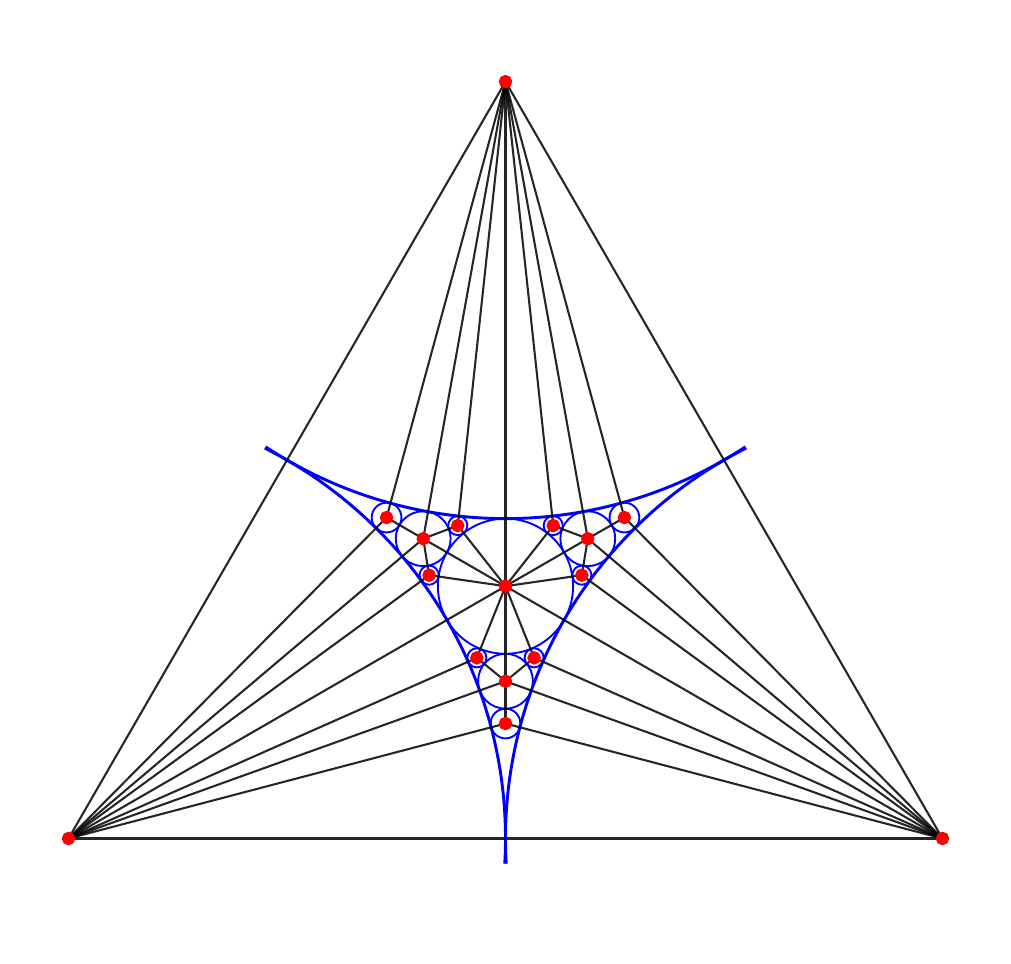}
    \caption{Third generation (\(n=3\)) of the Apollonian network embedded in the corresponding Apollonian circle packing. Nodes are located at the centers of tangent circles, and edges connect pairs of sites associated with mutually touching circles. New sites are recursively introduced inside triangular interstices and linked to the vertices of the enclosing triangle.}
    \label{fig:Rede_Apolo}
\end{figure}

The AN originates from the concept of the Apollonian packing, which comes as a solution to the problem of space-filling of spheres. In this solution, three circles touching each other creates a gap, which is filled by a new circle that is also tangent to the three original circles. This process generates three smaller gaps, each of which is subsequently filled with another circle, and this process is repeated, forming the Apollonian packing. In a given generation $n$, the AN is constructed by establishing connections between the centers of all pairs of circles that touch each other.
In Fig. \ref{fig:Rede_Apolo} we illustrate the third generation of the Apollonian network (\(n=3\)), embedded in the underlying Apollonian circle packing. Generalizations of the Apollonian network have been proposed, including binary and random variants, which preserve hierarchical growth while allowing for tunable structural and spectral properties such as clustering and localization \cite{meu1,meu2}.

ANs provide a particularly rich backbone to investigate a variety of physical models, such as 
quantum walks \cite{xu08}, coupled-cavity systems \cite{almeida13}, magnetism \cite{andrade05mag,kaplan09}, Bose-Einstein condensates \cite{oliveira10}, and electronic models \cite{souza07,cardoso08,oliveira09}.
Their hierarchical and self-similar properties imprints a gapped spectral structure featuring 
minibands, discrete modes, degenerate levels, and multifractality \cite{andrade05physa,oliveira09}. 
As a consequence,   
localization occurs naturally, in the absence of disorder, and strongly depends on the energy level and node degree \cite{cardoso08,almeida13}.  

From the standpoint of condensed-matter physics, 
Anderson localization is a cornerstone phenomenon in wave transport and arises as a consequence of disorder \cite{evers08}].
A single particle in the tight-binding approximation may be subjected to random on-site potentials and/or hopping amplitudes, thereby suppressing diffusion due to interference effects. 
%
Off-diagonal (hopping) disorder promotes distinct localization behavior compared to diagonal disorder, particularly in bipartite lattices
featuring sublattice symmetry \cite{Eilmes2001,Slanina2017}. 

Our goal here is to connect these disorder-induced mechanisms with the builtin localization properties of complex networks. From a general perspective, it is paramount to understand how topology competes with randomness in shaping localization patterns \cite{skipetrov22,sade05,jahnke08,goda06, zeng24, broni26}.
Previous studies on ANs have mainly relied on global indicators, such as the participation ratio, and focused on specific spectral regions \cite{cardoso08,almeida13}. 
While providing insight into the localization properties of the AN, these measures do not reveal where the modes localize. Moreover, metrics commonly used in regular lattices, based on spatial decay along a well-defined axis, are not directly applicable to complex networks. This limitation is particularly relevant for ANs, whose heterogeneous structure suggests that localization is strongly tied to node type (e.g., hub, corners, and low-degree sites).

In this work, we address this point by analyzing localization in the AN within a tight-binding framework, considering both diagonal (on-site potential) and off-diagonal (hopping) disorder. Rather than relying on global metrics, we examine the spatial distribution of each eigenstate by identifying the sites of maximum amplitude. This approach allows us to resolve how localization depends on energy, disorder strength, and node degree.

\section{METHODS}

To investigate the localization properties of the AN, we follow the usual tight-binding approach by defining the Hamiltonian
\begin{equation}
H =\sum_{i}\epsilon_{i}{a}^{\dagger}_{i}{a}_{i} + \sum_{\langle i,j \rangle} t_{ij}{a}^{\dagger}_{i}{a}_{j},
\label{eq1}
\end{equation}
where ${a}^{\dagger}_{i}$ (${a}_{i}$) is the single-particle creation (annihilation) operators at site $i$, $\epsilon_{i}$ represents its on-site potential, and $t_{ij}$ is the hopping strength between sites $i$ and $j$. The second sum runs over pairs of connected sites $\langle i,j \rangle$ as encoded by the adjacency matrix of the AN.
This Hamiltonian form can be used to investigate a variety of phenomena in different contexts \cite{almeida13,andrade05mag,souza07,cardoso08}.

In this work, we want to analyze the influence of uncorrelated diagonal and off-diagonal disorder on the Hamiltonian above. 
We do so by assigning each on-site energy $\epsilon_{i}$ and hopping amplitude $t_{ij}$ as independent random variables drawn
from the uniform intervals $[-\Delta /2,\Delta /2]$ and $[t-\delta /2,t+\delta /2]$, respectively, where 
$\Delta$ and $\delta$ are the disorder widths. From now on we express the energy in units of $t\equiv 1$.
In the simulations that follow we employ
$\Delta,\delta=0.01$, $1.00$, and $1.99$ to represent, respectively, the weak, intermediate, and strong disorder regimes.

Standard approaches to analyze the effects of disorder in regular lattices are often based upon testing localization measures against the system size. 
In our case, however, the growth of the AN is given by its generation $n$, rendering $N(n)=(3^n+5)/2$ sites \cite{prlApolo}, what rapidly imposes a computational limit. 
%
Fortunately, despite the nontrivial mode structure of the AN, its spectrum exhibits recurrence across generations \cite{andrade05physa} due to the self-similarity of the network associated with many layers of $C3$ symmetry. 
Consequently, localization is intrinsically dependent on energy, and we find that
$n=6$ ($N=367$ sites) already captures the main localization features.

The property we focus on to characterize localization in the AN is the maximum site probability amplitude at a given energy. More specifically, we numerically solve the eigenvalue 
problem $H\ket{E_\mu}=E_\mu\ket{E_\mu}$ via exact numerical diagonalization, where $\ket{E_\mu}$ are the eigenvectors and $E_\mu$ the corresponding energies. The site probability amplitude is then defined as $P_{\mu,i}=|\langle i|E_\mu\rangle |^2$, where $\ket{i}=a_i^{\dagger}\ket{\mathrm{vacuum}}$ are the local single-particle states. The maximum probability is then defined as 
$P_{\max,\mu}=\max_i \lbrace P_{\mu,i} \rbrace$ over $i$. 
Data are accumulated over all disorder realizations (we use 200) and grouped into small energy bins before averaging. This procedure effectively yields $P_{\max,\mu}\rightarrow P_{\max}(E)$.


\section{RESULTS}
\label{sec:sec3}




\begin{figure}[!h]
    \centering
    \includegraphics[width=1\linewidth]{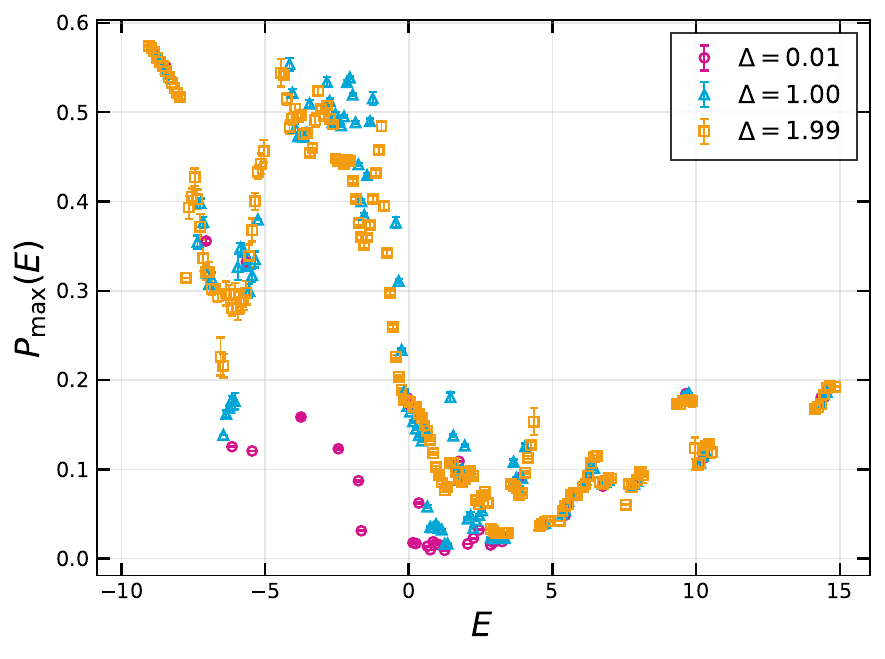}
    \caption{
    Maximum probability $P_{\max}$ as a function of the energy $E$ for different strengths of diagonal disorder $\Delta$.  
    Its increase has a noticeable effect for a range of negative energies. Vertical bars represent the standard error.}
    \label{fig:AndPmax}
\end{figure}

\begin{figure}[!h]
    \centering
    \includegraphics[width=1\linewidth]{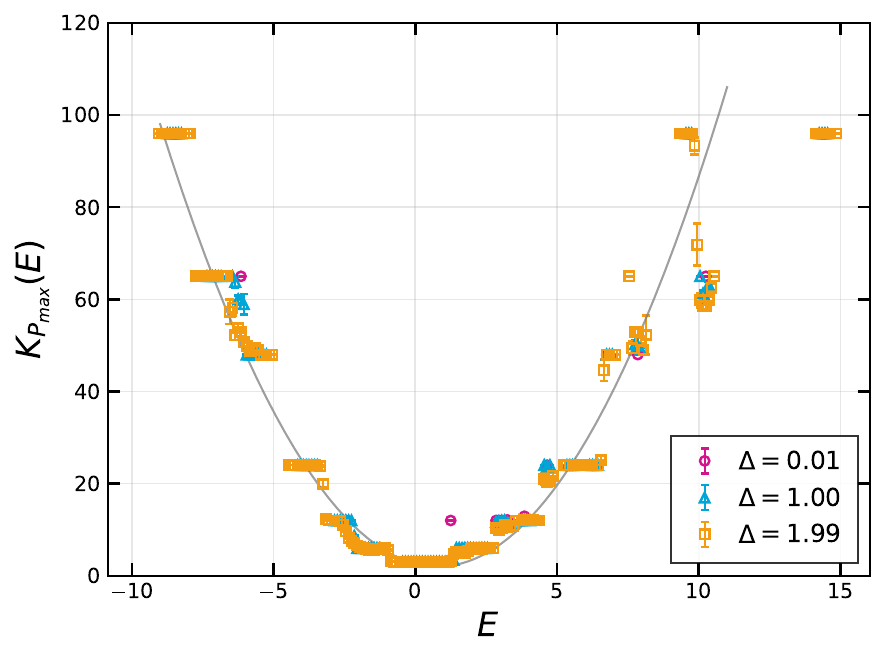}
    \caption{Node degree $K_{P_{\max}}(E)$ associated with the maximum probability $P_{\max}(E)$. Compare the symmetric degree distribution through the spectrum with the asymmetric localization profile seen in Fig. \ref{fig:K_pmax_and}. Vertical bars represent the standard error. The convex curve is a guide to the eye.}
    \label{fig:K_pmax_and}
\end{figure}

We first analyze localization in the AN under diagonal disorder. Figure \ref{fig:AndPmax} shows $P_{\max}$ across the energy spectrum for different disorder strengths $\Delta$.  
The negative end of the band supports strongly localized states regardless of the disorder level.
This is therefore a builtin property of the network, which arises due to the highly connected
central hub, as we will see shortly. As we move to the right, $P_{\max}$ experiences a slight drop and then 
reaches about the previous degree of localization. This time, however, it is clearly induced by the disorder upon increasing $\Delta$, featuring a moderate response at the band center ($E=0$). Its right side is mostly characterized by mild localization with 
negligible influence of the disorder. We remark that as the spectrum of the AN features many gaps and degenerate levels, there are fewer points in the weak disorder regime ($\Delta=0.01$) since fluctuations are reduced. The same reasoning applies to all the figures shown hereafter. 

To identify which network sites are responsible for the observed localization in different spectral regions, we next analyze the degree $k$ of the site associated with $P_{\max}$, defined as $K_{P_{\max}}$, as a function of energy. Results are shown in
Fig. \ref{fig:K_pmax_and}, where a clear separation emerges between spectral regions dominated by highly connected sites from earlier generations and those dominated by low-degree sites introduced at later stages of the network growth, especially near the band center. Specifically, there are 
$3^{n-1}, 3^{n-2}, 3^{n-3}, \ldots, 3^{2},3,1,3$, sites with degree 
$k = 3, 3\cdot 2, 3\cdot 2^{2}, \ldots, 3\cdot 2^{n-2},\, 3\cdot 2^{n-1},2^n+1$, respectively, with the last
two entries representing the central hub and the three corners (see Fig. \ref{fig:Rede_Apolo}). Note that the values shown in Fig. \ref{fig:K_pmax_and} are not restricted to these exact values of $k$ as the results are ensemble-averaged.  

Despite the asymmetry observed in Fig. \ref{fig:AndPmax}, with strongly localized states lying at
negative energies, the degree distribution over $E$ is nearly symmetric with respect to $E=0$, apart from
a few outliers on the far right side of the band, separated by a large gap.
This reflects the hierarchical organization of the AN, with the hub acting as dominant site at extreme eigenvalues \cite{cardoso08,meu1,meu2} 
In contrast, near $E=0$ such a role is taken over by the $3^{n-1}$ sites introduced in the current generation, namely those with $k=3$. This is remarkable because there
is a known $\frac{3^{n-1}-3}{2}$-fold degeneracy at zero energy in the clean case whose states have support on $k=3$ sites only \cite{andrade05physa}. Even though the diagonal disorder naturally lifts such degeneracy and increases the localization degree (cf. Fig. \ref{fig:AndPmax}), the dominance of lowest-degree sites is preserved. 

As the number of $k=3$ sites exceeds the number of available degenerate states at $E=0$ when $\Delta=0$, then weak localization is observed around this region in the case of weak disorder. 
On the other hand, as the disorder is increased  
large values of $P_{\max}\sim10^{-1}$ are obtained, exceeding by about two orders of magnitude that of a reference state evenly distributed over the $k=3$ subset formed by 243 sites. 
The onset of localization around $E=0$ reveals that 
the break of the $C3$ symmetry favors particular spatial configurations while keeping the influence of low-degree nodes.


\begin{figure}[!h]
    \centering
    \includegraphics[width=1\linewidth]{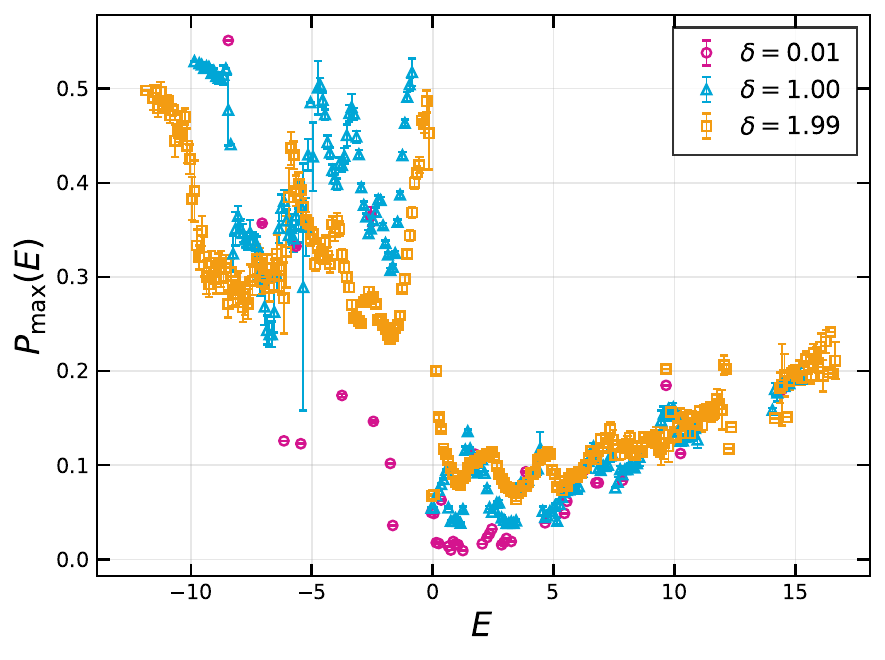}
    \caption{Maximum probability $P_{\max}$ as a function of the energy $E$ for different strengths of off-diagonal disorder $\delta$. Vertical bars represent the standard error.}
    \label{fig:ligPmax}
\end{figure}

\begin{figure}[!h]
    \centering
    \includegraphics[width=1\linewidth]{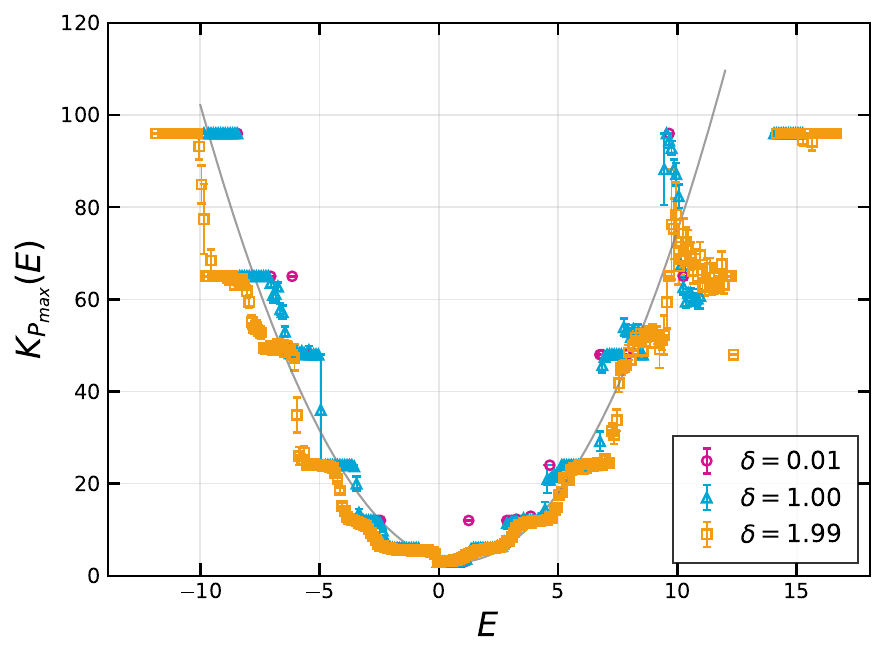}
    \caption{Node degree $K_{P_{\max}}(E)$ associated with the maximum probability $P_{\max}(E)$ in the presence of off-diagonal disorder with different strengths $\delta$. Vertical bars represent the standard error. Convex curve is for guiding the eye. The convex curve is a guide to the eye.}
    \label{fig:K_pmax_lig}
\end{figure}

We now move on to the effects of off-diagonal (hopping) disorder on the AN. In this case, randomness is directly introduced in the hopping amplitudes $t_{ij}$, thereby affecting
the network structure.
Figures \ref{fig:ligPmax} and \ref{fig:K_pmax_lig} shows $P_{\max}(E)$ 
and their associated $K_{P_{\max}}(E)$ for different disorder strengths $\delta$.
For the most part, their behavior are qualitatively similar to the previous case of diagonal disorder.  
However, some differences emerge in the negative part of the spectrum, where
$P_{\max}(E)$ exhibits a more unstable behavior. Most importantly, we observe that the localization weakens as $\delta$ increases from $1.00$ to $1.99$. This is also manifested, albeit  
to a lesser extent, in Fig. \ref{fig:AndPmax}.

Another point of interest is that the degeneracy present at $E=0$ is maintained, now with one state less.
Although off-diagonal disorder breaks the underlying $C3$
symmetry of the clean AN, 
the zero modes originate from local interference constraints and exist independently of the 
$C3$ symmetry. The latter only enforces equivalence between the self-similar triangular branches of the AN, leading to an enhanced degeneracy with a particular set of compact eigenstates.
This issue is investigated in more detail elsewhere \cite{souza26flat}. Here, we mention that 
the eigenstates associated with $E=0$ and its vicinity become more localized, indicating that their spatial reconfiguration is also a consequence of the symmetry breaking, which affects low-degree nodes more strongly, as in the case of diagonal disorder.


\begin{figure}[!t]
    \centering
    \includegraphics[width=1\linewidth]{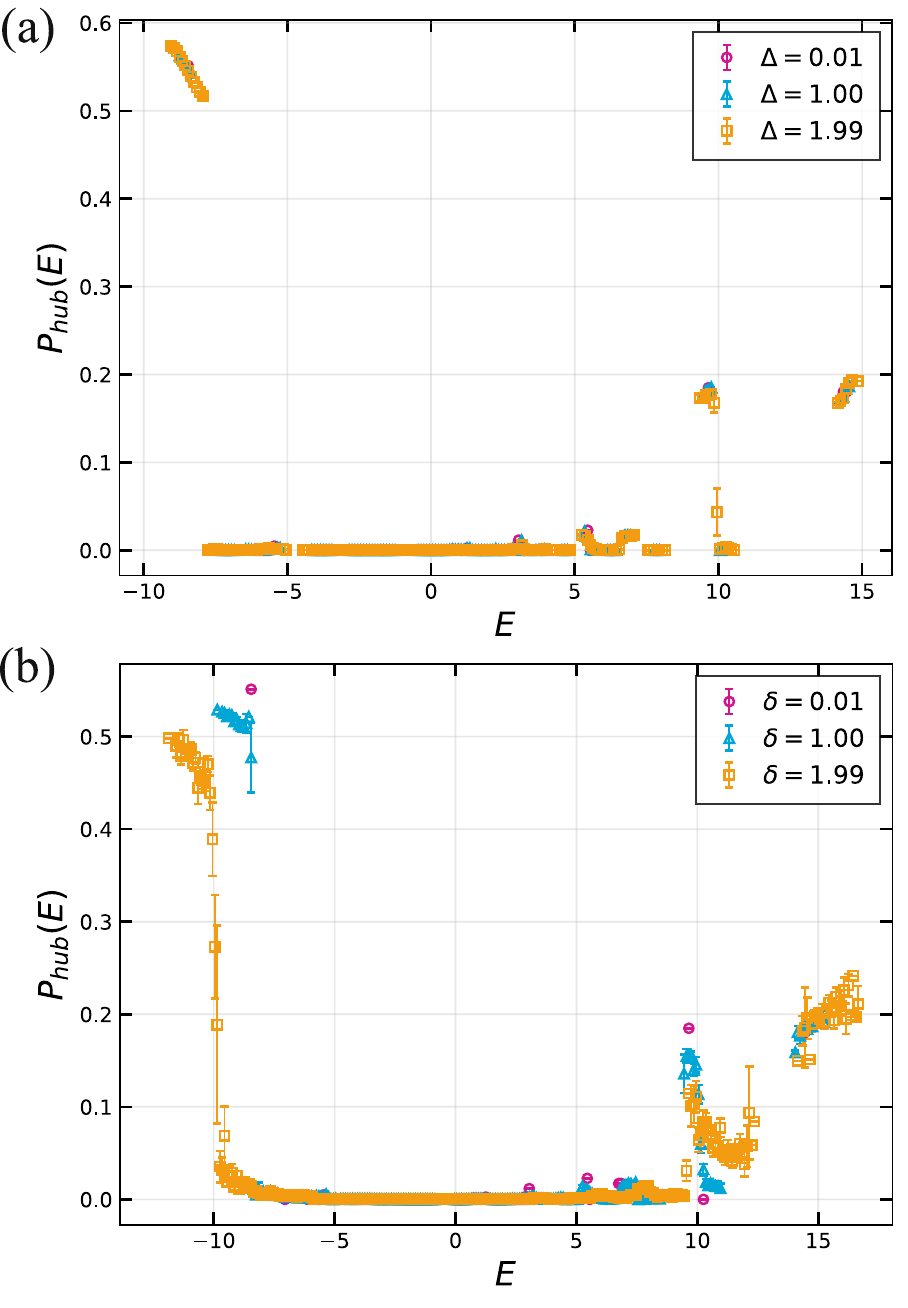}
    \caption{Hub probability amplitude $P_{\mathrm{hub}}$ as a function of energy $E$ in the presence of (a) diagonal and (b) off-diagonal disorder, for different associated disorder strengths $\Delta$ and $\delta$, respectively. Vertical bars represent the standard error.}
    \label{fig6}
\end{figure}

In this final part of the discussion, we focus on the role of the hub by tracking 
$P_{\mathrm{hub}}$ along the spectrum. Results are shown in Fig. \ref{fig6} for both diagonal and off-diagonal disorder.
Again, we confirm their influence at both ends of the spectrum, with stronger support on the extreme left. Now, we also observe that the hub has essentially negligible support elsewhere, spanning a significant region around the band center, even in the presence of different types of perturbations.
This confirms that the builtin localized states of the AN, arising from the high connectivity of the hub, are robust against imperfections.

\section{CONCLUSIONS}
\label{sec:sec4}

In this work, we have analyzed localization in the AN under diagonal and off-diagonal disorder using a site-resolved measure. We showed that localization is strongly energy-dependent and closely tied to node degree. Eigenstates at the spectral edges are dominated by the hub while those near the band center are associated with low-degree nodes, for any disorder strength. This reflects the robustness of the underlying hierarchical organization of the network. While both disorder mechanisms yield qualitatively similar behavior, off-diagonal disorder introduces notable differences, including weakening of localization induced by disorder
at negative energies and the persistence of degeneracy at zero energy.

Although our results were presented for a single generation ($n=6$), we have verified that the localization patterns persist across different generations at fixed disorder strength. This indicates that the observed behavior is not a finite-size effect, but rather a direct consequence of the self-similar and hierarchical structure of the AN.

Despite its synthetic construction, the AN shares properties commonly observed in real systems, as it is scale-free and displays small-world behavior. In this context, 
our findings establish the AN as a paradigmatic platform in which disorder and network geometry interplay to produce rich and highly structured localization phenomena. These results provide guidance for future studies of quantum transport in hierarchical and fractal networks \cite{xu08,almeida13}. Further investigations could address localization in generalized variants of the AN \cite{meu1,meu2}, as well as the subtle localization properties of degenerate levels under symmetry breaking \cite{souza26flat,broni26}.

\begin{acknowledgments}
We thank A. M. C. Souza for sharing key insights and fruitful
discussions. This work was supported by CNPq and CAPES.
\end{acknowledgments}



%

\end{document}